\documentclass[useAMS,usenatbib]{mn2filippo}
\usepackage{epsfig}
\usepackage{deluxetable} 
\usepackage{graphicx,amssymb,amsmath,times}

\begin{document}

\title[Long-term monitoring of PKS 0537--441 with {\em Fermi}-LAT and multiwavelength observations]{Long-term monitoring of PKS 0537--441 with {\em Fermi}-LAT and multiwavelength observations}
\author[D'Ammando, Antolini, Tosti, et al.]{F. D'Ammando$^{1,2}$\thanks{E-mail: filippo.dammando@fisica.unipg.it}, E. Antolini$^{1,2}$, G. Tosti$^{1,2}$, J. Finke$^{3}$, S. Ciprini$^{4}$, S. Larsson$^{5, 6, 7}$, \newauthor M. Ajello$^{8,9}$, S. Covino$^{10}$, D. Gasparrini$^{4}$, M. Gurwell$^{11}$, M. Hauser$^{12}$, P. Romano$^{13}$, \newauthor F. Schinzel$^{14}$, S. J. Wagner$^{12}$, D. Impiombato$^{13}$, M. Perri$^{4}$, M. Persic$^{15,16}$, E. Pian$^{15,17}$, \newauthor G. Polenta$^{4,18}$, B. Sbarufatti$^{9,19}$, A. Treves$^{20}$, S. Vercellone$^{13}$, A. Wehrle$^{21}$, A. Zook$^{22}$\\
$^{1}$Dipartimento di Fisica, Universit\`a degli Studi di Perugia, I-06123 Perugia, Italy \\
$^{2}$Istituto Nazionale di Fisica Nucleare, Sezione di Perugia, I-06123 Perugia, Italy \\ 
$^{3}$U.S. Naval Research Laboratory, Code 7653, 4555 Overlook Ave. SW, Washington, DC 20375-5352, USA \\
$^{4}$Agenzia Spaziale Italiana (ASI) Science Data Center, I-00044 Frascati (Roma), Italy\\
$^{5}$Department of Physics, Stockholm University, Albanova, SE-106 91 Stockholm, Sweden \\
$^{6}$The Oskar Klein Centre for Cosmoparticle Physics, Albanova, SE-106 91 Stockholm, Sweden \\
$^{7}$Department of Astronomy, Stockholm University, SE-106 91 Stockholm, Sweden \\ 
$^{8}$Kavli Institute for Particle Astrophysics and Cosmology, Dep. of Physics and SLAC National Accelerator Laboratory, Stanford University, Stanford, CA 94305, USA \\   
$^{9}$Space Sciences Laboratory, 7 Gauss Way, University of California, Berkeley, CA 94720- 7450, USA \\
$^{10}$INAF - Oss. Astronomico di Brera, Merate (LC), I-23807, Italy \\
$^{11}$Harvard-Smithsonian Center for Astrophysics, Cambridge, MA 02138, USA \\
$^{12}$Landerssternwarte, Universit\"at Heidelberg, K\"onigstuhl,
  D-69117 Heidelberg, Germany \\
$^{13}$INAF-IASF Palermo, Via Ugo La Malfa 143, Palermo, I-90146, Italy \\
$^{14}$Department of Physics and Astronomy, University of New Mexico, MSC07 4220, Albuquerque, NM 87131-0001, USA \\
$^{15}$INAF-OA Trieste, Via G. Tiepolo 11, I-34127 Trieste, Italy \\
$^{16}$INFN, Sezione di Trieste, I-34127 Trieste, Italy \\
$^{17}$Scuola Normale Superiore di Pisa, Piazza dei Cavalieri 7, I-56126 Pisa, Italy \\
$^{18}$INAF - Oss. Astronomico di Roma, Via Frascati 33, I-00040 Monteporzio Catone, Italy \\
$^{19}$Department of Astronomy and Astrophysics, Pennsylvania State University, University Park, PA 16802 USA \\
$^{20}$Dep. of Physics and Mathematics, University of Insubria, I-22100 Como, Italy \\
$^{21}$Space Science Institute, Boulder, CO 80301, USA \\
$^{22}$Jet Propulsion Laboratory, California Institute of Technology, CA 91109 Pasadena, USA}

\maketitle

\begin{abstract}

We report on multiwavelength observations of the blazar PKS 0537$-$441 ($z$ =
0.896) obtained from microwaves through $\gamma$ rays by SMA, REM, ATOM, \textit{Swift} and
\textit{Fermi} mostly during 2008 August--2010 April. Strong variability has been observed in $\gamma$ rays, with two major flaring episodes (2009 July and 2010 March) and a
harder-when-brighter behaviour, quite common for flat spectrum radio
quasars and low-synchrotron-peaked BL Lacs, in 2010 March. In the same way the spectral energy distribution (SED) of the source cannot be modelled by a simple
synchrotron self-Compton model, as opposed to many BL Lacertae objects, but the addition of
an external Compton component of seed photons from a dust torus is needed. The 230 GHz light curve showed an increase
simultaneous with the $\gamma$-ray one, indicating co-spatiality of the mm and $\gamma$-ray
emission region likely at large distance from the central engine. The low, average, and high
activity SED of the source could be fit changing only the electron
distribution parameters, but two breaks in the electron distribution are
necessary. The ensuing extra spectral break, located at NIR-optical frequencies, together with that in $\gamma$ rays seem to indicate a common origin, most likely due to an intrinsic feature in the underlying electron distribution.
An overall correlation between the $\gamma$-ray band with the $R$-band and
$K$-band has been observed with no significant time lag. On the other hand,
when inspecting the light curves on short time scales some differences are
evident. In particular, flaring activity has been detected in NIR and optical bands with no evident $\gamma$-ray counterparts in 2009 September and November. Moderate variability has been observed in X-rays
with no correlation between flux and photon index. An increase of the
detected X-ray flux with no counterpart at the other wavelengths has been
observed in 2008 October, suggesting once more a complex correlation between the
emission at different energy bands.

\end{abstract}


\section{Introduction}\label{sect:Intro}

PKS~0537$-$441 (also known as HB89~0537$-$441, RX~J0538.8$-$4405) was selected as a quasi-stellar source in the Parkes 2700 MHz
radio survey \citep{Peterson1976} and identified as a blazar by \citet{Burbidge1992}. According to its spectral energy distribution the source was classified
as a BL Lac object, categorized as a low-energy-peaked BL Lac by \citet{padovani95}, although
the determination of the redshift of this source is based on broad emission
lines and the object has been listed as a highly polarized quasar
\citep{Ledden1985, sambruna94}, having radio polarization up to 3$\%$ \citep{komesaroff84}
and optical polarization up to 18$\%$ \citep{impey88, impey90}. PKS 0537$-$441 has also been suggested to be a transition object between classical
BL Lac objects and violently variable and highly polarized quasars
\citep{cristiani85, maraschi85, treves93, giommi95, ghisellini11}.

Two broad emission lines at 3617\AA\, and 5304\AA\, were observed
in the optical spectrum of PKS~0537$-$441, and proposed to correspond to the C
III$]$ $\lambda$1909 and MgII $\lambda$2798 lines, placing the object at a redshift z = 0.894
  \citep{Peterson1976, Wilkes1983, Stickel1993}. Subsequently,
  \citet{Lewis2000} reported a redshift of z = 0.892 $\pm$ 0.001,
  based on the Balmer line series and the [O III] emission line. PKS~0537$-$441 was also observed by the Faint Object Spectrograph (FOS) of the \textit{Hubble Space
Telescope} (HST) in 1993 July and September and, on the basis of their measured central wavelengths, the emission lines detected in the
HST FOS spectra were identified by \citet{Pian2005} with Ly $\alpha$, Si IV and C IV at z =
0.896 $\pm$ 0.001, a value consistent with the original redshift
determination by \citet{Peterson1976}. 
Based on a claimed detection of a galaxy along the line of sight to PKS
0537$-$441, it was discussed as a case of gravitational lensing
\citep{Stickel1988, Lewis2000}, in which a foreground galaxy lenses the
radiation from the quasar, amplifying the continuum with respect to the
line emission, but further observations reported in \citet{falomo92, Pian2002, heidt03} have not confirmed the presence of a foreground galaxy. 

Given its apparent brightness, PKS 0537$-$441 is an excellent target for
studying the properties of blazars and therefore it has been the subject of several monitoring campaigns from radio to
optical, showing remarkable variability. \citet{peterson72}, who found Quasi Stellar Objects (QSOs)
among the Parkes radio sources, discovered the optical counterpart of the
radio source PKS 0537$-$441 to be 2 mag brighter than on the Palomar Observatory Sky
Survey. Subsequent observations by \citet{eggen73} showed a variation between
magnitude 16.5 and 13.7 over a few months.
The study of the Harvard Observatory photographic plate collection \citep{liller74} over almost a century revealed
long-term variations of $\sim$5 mag with fluctuations of $\sim$2 mag over less
than 2 months. In addition this source displayed intraday variability at
radio \citep{romero95}, and optical frequencies \citep{heidt96, romero02}. Although PKS 0537$-$441 is one of the brightest radio selected BL
Lac objects \citep{stickel91} it is a relatively weak X-ray source
\citep{tanzi86, worral90}, even if the comparison of the ROSAT X-ray spectrum with
the \textit{Einstein} and EXOSAT observations indicates substantial
variability both in intensity and spectral slope at
X-rays \citep[and the reference therein]{treves93}. BeppoSAX in November 1998 observed the
source with a spectrum and emission state consistent with
those measured by EXOSAT and \textit{Einstein}, but with a flux at 1 keV almost
a factor of two less than that detected by ROSAT in 1991. Moreover, the
BeppoSAX spectrum suggests that a single emission component dominates in the energy range 0.1--30 keV \citep{Pian2002}. 
 
Previous multifrequency observations from near-IR to
X-rays \citep{maraschi85, tanzi86} showed quasi-simultaneous flares at IR, optical, and X-ray frequencies.
Its continuum variations throughout the entire electromagnetic spectrum have
been discussed in detail by \citet{Pian2002}.  
In 2005 PKS 0537--441 was monitored in the optical and
infrared by REM \citep{dolcini05} and observed by all the instruments on
board \textit{Swift} in January, July, and November \citep{pian07}, with a flux that
varied by a factor of $\sim$60 and $\sim$4 in optical and X-rays, respectively. The
$V$-band and X-ray light curves measured in 2004--2005 were highly correlated,
although with different variation amplitudes. In contrast, no clear evidence of
variability within a single night was found, even though some hints of small flares
on day time-scales may be present. Moreover the X-ray photon index was
observed to be steady over the different epochs, as opposed to the variation
of the fluxes. The optical spectra collected at different epochs suggest the presence
of thermal emission during the low states. Only recently,
\citet{impiombato11} identified an episode of rapid variability in the $J$-band with a duration of
$\sim$ 25 minutes.

PKS 0537--441 was detected in $\gamma$ rays as 3EG J0540--4402 by the EGRET telescope on board the \textit{Compton Gamma-Ray Observatory}
\citep{hartman99}. The source was detected by EGRET for the first time in 1991
\citep{michelson92, thompson93} and then re-observed at many
successive epochs in different states \citep[see also][]{treves93}, showing
bright and variable emission in $\gamma$ rays,
with a maximum flux of 9$\times$10$^{-7}$ photons cm$^{-2}$ s$^{-1}$ during
1995 January and a peak with the temporal binning of 2 days of (20$\pm$5)$\times$10$^{-7}$ photons cm$^{-2}$ s$^{-1}$ \citep{Pian2002}. More recently intense $\gamma$-ray activity from this source was
observed by the Large Area Telescope (LAT) onboard \textit{Fermi}
\citep{tosti08, bastieri09, cannon10} and AGILE Gamma-Ray Imaging Detector \citep{lucarelli10}. PKS
0537$-$441 was listed in both the First and Second \textit{Fermi} Large Area Telescope Source catalogs \citep{Abdo2010b, nolan12}.

Now the \textit{Fermi} satellite, with continuous monitoring of the entire $\gamma$-ray sky and the excellent
sensitivity of the LAT, gives us the opportunity to study the blazars in the $\gamma$-ray band not only during
flaring states but also during low activity states, broadening correlated investigations of blazars over the whole
electromagnetic spectrum. In this context the \textit{Swift} satellite, with
broadband coverage and scheduling flexibility create a perfect synergy with
\textit{Fermi} and provide crucial data in the gap in coverage between the $\gamma$-ray data and the radio-to-optical data obtained from ground-based telescopes allowing a deep and complete long-term monitoring of these sources.

In this paper we report the results of the LAT monitoring of PKS 0537--441
together with the related multiwavelength observations across the electromagnetic
spectrum. This paper is organized as follows. In Section 2 we present LAT data
and analysis. In Section 3 we present the multifrequency data collected by {\it Swift}, REM,
ATOM, and SMA. In Section 4 we discuss the $\gamma$-ray spectral and flux
variability, while the correlation with the other energy bands is discussed in
Section 5. The general
properties of the source and a comparison of the multifrequency data reported
in this paper with respect to past observations are discussed in Section 6. In Section 7 we
discuss the SED modeling for different epochs. Finally in Section 8 we draw our conclusions.

In the following we use a $\Lambda$CDM (concordance) cosmology with values given within 1$\sigma$ of the Wilkinson Microwave Anisotropy Probe (WMAP)
results \citep{komatsu09}, namely \textit{h} = 0.71, $\Omega_m$ = 0.27, and
$\Omega_{\Lambda}$ = 0.73, and a Hubble constant value \textit{H$_0$} = 100
\textit{h} km s$^{-1}$ Mpc$^{-1}$, with the corresponding luminosity distance $d_L \simeq 5.78$\ Gpc ($\sim 1.8\times10^{28}$\ cm).

\section{LAT observations}\label{sect:LAT_data_analaysis}

\par  The {\em Fermi}-LAT is a pair-conversion $\gamma$-ray detector, sensitive to photon energies from about 20 MeV to $>$300 GeV. It consists
of a tracker (composed of two sections, front and back, with different capabilities), a calorimeter and an anticoincidence system to reject the
charged-particle background. The LAT has a large peak  effective area  ($\sim 8000$~cm$^2$ for 1 GeV photons in the event class considered here), viewing
$\approx 2.4$~sr of the sky with single-photon angular resolution (68\% containment radius) of $0.6^\circ$ at $E = 1$~GeV on-axis \citep{Atwood09}. 

\par The data presented in this paper were collected in the first 20 months of
     {\it Fermi} science operation,
from 2008 August 4 to 2010 April 4 ($\sim$ 600 days, from MJD 54682 to 55291).
     The analysis was performed with the standard {\em Fermi} LAT
     \texttt{ScienceTools} software
     package\footnote{http://fermi.gsfc.nasa.gov/ssc/data/analysis/documentation/Cicerone/}
     (version v9r15p6). Only events having the highest probability of being photons, belonging to the ``Diffuse'' class, in
     the energy range 0.1--100 GeV were used in the analysis. The instrument
     response functions (IRFs) P6\_V3\_DIFFUSE were used.
In order to avoid significant background contamination from Earth limb $\gamma$ rays, all events with
zenith angle $> 105^{\circ}$ were excluded. 
In addition only photons detected when the spacecraft rocking angle was $< 52^{\circ}$ were
selected. This eliminated time intervals with the Earth in the LAT field of view.

\begin{figure*}
\centering
\includegraphics[width=16.5cm,angle=-90]{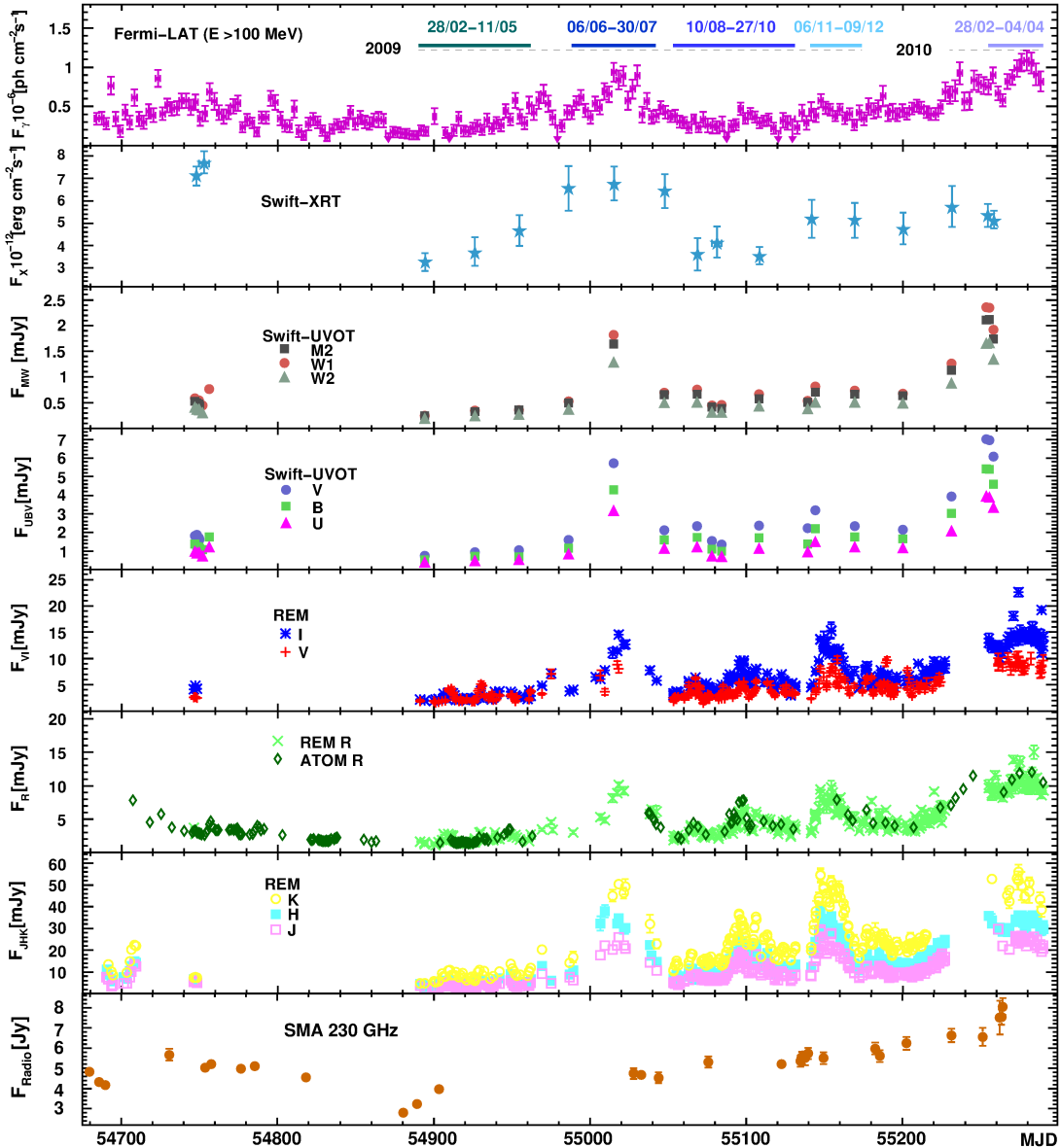}
\caption{Multifrequency light curve for PKS
  0537--441. The period covered is  2008 August 4 -- 2010 April 4.
  The data sets were collected (from top to  bottom) by {\it Fermi}-LAT
  ($\gamma$ rays), {\it Swift}-XRT (0.3--10 keV), {\it Swift}-UVOT ($W1$, $M2$, $W2$, and $V$, $B$, $U$ filters), REM ($V$ and $I$ bands),
  REM and ATOM ($R$ band), REM ($J$, $H$, $K$ bands), and SMA (230 GHz). The
  solid lines in the top panel represent the time intervals used in the LAT spectra extraction. 
}
\label{FIG:MW_lc}
\end{figure*}

The LAT light curves were computed using the maximum-likelihood algorithm
implemented in \textit{gtlike}. The 20-month light curve was built using
3-day time bins (Fig.~\ref{FIG:MW_lc}, upper panel); for each time bin integrated
flux, photon index, and test statistic (TS)\footnote{The test statistic (Mattox
  et al. 1996) is defined as TS\,= $-2(\log L_0 - \log L)$, where $L_0$ is the
  likelihood for the null hypothesis (i.e. no source exists at
the given position) and $L_1$ is the alternative hypothesis (the source exists).} values were determined.
For each time bin we selected photons included in a region of interest (RoI) of $12^{\circ}$ in radius, centered on the position of the source. 
In the RoI analysis the source is modelled with a power-law, $dN/dE = (N(\Gamma +1) E^{-\Gamma})/(E_{max}^{\Gamma+1}-E_{min}^{\Gamma+1})$, where $N$
is the normalisation, $\Gamma$ is the photon index, E$_{min}$ and E$_{max}$
are the limits of the energy interval chosen for the Likelihood analysis.

All point sources listed in the First Fermi LAT catalog \citep[1FGL;][]{Abdo2010a} within $19^{\circ}$
from PKS 0537--441 with TS $>50$ and fluxes above 10$^{-8}$ ph
cm$^{-2}$s$^{-1}$ were included in the RoI model using a power-law
spectrum. The background model used to extract the $\gamma$-ray signal includes
a Galactic diffuse emission component and an isotropic component. The Galactic
component is parameterized by the map cube file
gll\_iem\_v02.fit. The isotropic background component, which is the sum of residual instrumental background and
extragalactic diffuse $\gamma$-ray background, was included in the ROI model using the standard
model file isotropic\_iem\_v02.txt\footnote{http://fermi.gsfc.nasa.gov/ssc/data/access/lat/BackgroundModels.html}. 
In the light curve computation the photon index value was frozen to the
value obtained from the likelihood analysis over the entire period. For each
time bin, if the TS value for the source was less than 4 or the number of model predicted
photons \textit{N$_{pred}$} $<10$, a 2 $\sigma$ upper limit was computed for the flux. The
estimated relative systematic uncertainty on the $\gamma$-ray flux, which according to
\citet{Abdo2010b} reflects the relative systematic uncertainty on
effective area, is 10\% at 100 MeV, 5\% at 500 MeV, and 20\% at 10 GeV. All errors reported throughout the paper are statistical only.

\section{The multifrequency coverage}

The monitoring of PKS 0537--441 by {\em Swift} ranged from  optical to X-ray bands and was supplemented by data collected in near-IR 
and optical bands by REM and ATOM, as well as sub-mm by SMA. This provided excellent broadband coverage.

\subsection{{\it Swift}}

The {\it Swift} satellite \citep{gehrels04} performed several observations
of PKS 0537--441 between 2008 October 7 and 2010 March 4, with all three on board
experiments: the X-ray Telescope \citep[XRT;][]{burrows05}, the Ultraviolet/Optical Telescope \citep[UVOT;][]{roming05} and the coded-mask Burst
Alert Telescope \citep[BAT;][]{barthelmy05}. These observations were obtained
both by means of dedicated Target of Opportunities (ToOs) and by activating a monthly monitoring program
(PI: F. D'Ammando) covering the period 2009 March--2010 February.

\subsubsection{{\it Swift}/BAT}

The Burst-Alert Telescope on board the {\it Swift} satellite is a coded-mask
telescope operating in the 15--200 keV energy range. Thanks to its large field
of view, BAT surveys up to 80$\%$ of the sky every day. We selected all
observations with PKS 0537$-$441 in the BAT field of
view, between 2004 November and 2009 August. The data were
processed using {\tt Heasoft} package (v.6.8) and following the recipes presented in \citet{ajello09}. The spectrum of PKS 0537$-$441 was extracted using the method
presented in \citet{ajello08}.

\subsubsection{{\it Swift}/XRT}

The XRT data were processed with standard procedures ({\tt xrtpipeline}
v0.12.6), filtering, and screening criteria by using the {\tt Heasoft} package (v.6.11).
The source count rate was low during the whole campaign
(mean count rate $<0.5$ counts s$^{-1}$), so we considered only photon
counting (PC) data and further selected XRT event grades 0--12. 
Pile-up correction was not required.
Source events were extracted from a circular region with 20-pixel radius (1
pixel $\sim 2\farcs36$), while background events were extracted from an
annular region centered on the source and with radii of 55 and 95 pixels.
Ancillary response files were generated with {\tt xrtmkarf}, and account for 
different extraction regions, vignetting and PSF corrections. 

A spectrum was extracted from each observation and fit with
{\tt XSPEC} (v12.7.0) adopting an absorbed power law model with free photon index 
$\Gamma_{\rm X}$ and using the photoelectric absorption model {\tt tbabs}
with a neutral hydrogen column fixed to its Galactic value \citep[$2.9  \times
10^{20}$ cm$^{-2}$;][]{murphy96}, consistent with \citet{pian07}.
Data were rebinned to have at least 20 counts per energy bin to allow the $\chi^{2}$ minimization.

\subsubsection{{\it Swift}/UVOT}

During the {\it Swift} pointings, the UVOT \citep{poole08} instrument observed
PKS 0537$-$441 in the $V$, $B$, $U$, and $W1$, $M2$ and $W2$ photometric bands. 
The analysis was performed using the {\tt uvotsource} tool to extract counts from a standard 5\arcsec\, radius source aperture, correct for coincidence losses,
apply background subtraction, and calculate the source flux. The background
counts were derived from a circular region of 10\arcsec\, radius in the source
neighbourhood. The fluxes were then de-reddened using the values of E(B-V)
taken from \citet{schlegel98} with A($\lambda$)/E(B-V) ratios calculated for
the UVOT filters using the mean interstellar extinction curve from
\citet{fitzpatrick99}.

\subsection{REM}

The Rapid Eye Mounting \citep[REM;][]{zerbi01, covino04} is a robotic telescope located at the ESO Cerro La Silla observatory
(Chile). The REM telescope has a Ritchey-Chretien configuration with
a 60-cm f/2.2 primary and an overall f/8 focal ratio in a fast moving
alt-azimuth mount providing two stable Nasmyth focal stations. At one of the
two foci, the telescope simultaneously feeds, by means of a dichroic, two
cameras: REMIR for the NIR \citep{conconi04} and ROSS \citep{tosti04} for the
optical. The cameras both have a field of view of 10 $\times$ 10 arcmin and imaging capabilities with the usual NIR (z, $J$, $H$, and $K$) and Johnson-Cousins $VRI$ filters.
The REM software system  \citep{covino04} is able to manage complex
observational strategies in a fully autonomous way. All raw optical/NIR frames
obtained with REM telescopes were reduced following standard procedures. Instrumental magnitudes were obtained via aperture photometry and
absolute calibration has been performed by means of secondary standard stars
in the field reported in \citet{hamuy89} or by 2MASS\footnote{http://www.ipac.caltech.edu/2mass/} objects in the field.
The data presented here were obtained during 2008 August--2010 April by a
Guest Observer programme for announcement of observing time AOT18 (PI: D. Impiombato) and AOT19 (PI: F. D'Ammando) and a long term project for AOT20 and AOT21 (PI: F. D'Ammando).

\subsection{ATOM}

Optical observations in the Johnson $R$ filter for this campaign were
obtained between 2008 August and 2010 April with the 0.8-m optical
telescope ATOM in Namibia \citep{hauser04}. ATOM is operated robotically by the High Energy Stereoscopic System (H.E.S.S.) collaboration and obtains automatic observations of confirmed
or potentially $\gamma$-bright blazars. Data analysis \citep[debiassing, flat
fielding, and photometry with Source-Extractor,][]{bertin96} is conducted
automatically using our own pipeline. For differential photometry, the
reference stars 2, 3, and 6 from \citet{hamuy89} were used.

\subsection{SMA}

The 230 GHz (1.3 mm) light curve was obtained at the Submillimeter Array (SMA)
on Mauna Kea (Hawaii) from 2008 August 1 to 2010 March 9. PKS 0537--441 is included in an ongoing monitoring programme at the SMA to
determine the fluxes of compact extragalactic radio sources that can be used
as calibrators at mm wavelengths. Details of the observations and data
reduction can be found in \citet{gurwell07}. Data from this programme are
updated regularly and are available at SMA
website\footnote{http://sma1.sma.hawaii.edu/callist/callist.html. Use in
publication requires obtaining permission in advance.}. 
Additional SMA data are from a programme led by A. Wehrle\footnote 
{awehrle@spacescience.org} to monitor fluxes of blazars on the {\it Fermi} LAT Monitored
Source List\footnote{http://fermi.gsfc.nasa.gov/ssc/data/policy/LAT$\_$Monitored$\_$Sources.html}.

\section{Gamma-ray spectral and temporal variability}\label{spectral_variability}

\subsection{Spectral behaviour}

\begin{figure}
\centering
\includegraphics[width=8.5cm]{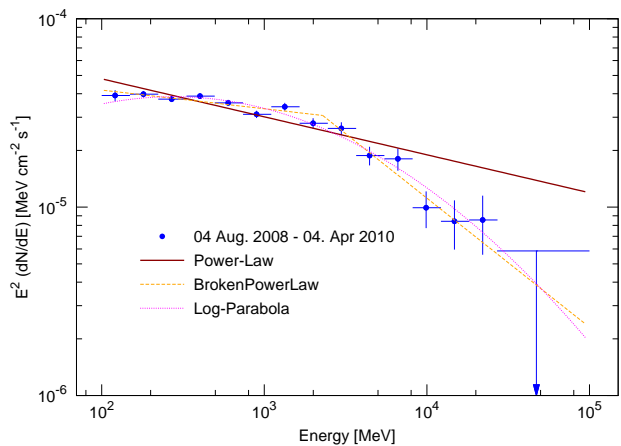}
\caption{20-month $\gamma$-ray spectrum of PKS 0537--441
 measured by {\it Fermi}-LAT between 2008 August 4 and 2010 April 4. 
 Solid brown line: power-law spectrum fit. Dashed orange line: broken power-law spectrum fit. Dotted magenta line: log-parabola spectrum fit.} 
\label{FIG:Fermispectrum20}
\end{figure}

The $\gamma$-ray spectral analysis of PKS 0537--441 was performed both for the first
20-month (2008 August 4 -- 2010 April 4; MJD 54682--55290) of {\em Fermi}-LAT observations, and
for 5 sub-periods: period 1 (2009 February 28--May 11; MJD 54890--54962),
period 2 (2009 June 6--July 30; MJD 54988--55042), period 3 (2009 August 10--October 27; MJD 55053--55131), period 4 (2009 November
6--December 9; MJD 55141--55174), and period 5 (2010 February 28--April 4; MJD
55255--55290). The sub-periods, reported also in Fig.~\ref{FIG:MW_lc}, have
been chosen for selecting different activity states in $\gamma$ rays. The results are reported
in Table~\ref{table:latresult}. A hardening of the $\gamma$-ray spectrum has been observed during period 5, which corresponds to the highest activity state. This ``harder when brighter'' behaviour was already reported in other bright flat spectrum radio quasars (FSRQs) and low-synchrotron-peaked (LSP) BL Lacs \citep{Abdo2010d}, even if only moderate variation ($\Delta\Gamma <$ 0.3) has been observed. By contrast we noted that during period 2 no hardening of the spectrum of PKS 0537$-$441 has been observed despite the high flux, indicating that the ``harder when brighter'' effect is not detected in all high activity periods of the source. The difference could be due to different causes or locations of the different flares. We noted that a ``harder when brighter'' trend is not a universal behaviour in blazar $\gamma$-ray flares. For example, in 3C 454.3, there is only very weak evidence of a ``harder when brighter'' trend \citep{ackermann10_3c454.3, abdo11_3c454.3} and there is no evidence for significant correlation between $\gamma$-ray flux and photon index in 3C 279 \citep{hayashida12}.
The 20-month spectrum is shown in Fig.~\ref{FIG:Fermispectrum20}. 
The energy spectrum was built by dividing the whole energy range (0.1--100 GeV)
into bands, requiring TS $>$ 50 and/or more than 8 photons predicted by the maximum likelihood source for each bin except for the last one. This results in 14 energy bins and
one upper limit for the 20-month spectrum, and 3 bins and one upper limit for
the 5 spectra in the sub-periods. For each energy bin a maximum likelihood analysis, fixing the spectral index at the respective global likelihood values computed
in the entire 0.1--100 GeV energy range, was performed.

By visual inspection the 20-month spectrum seems to be curved (Fig.~\ref{FIG:Fermispectrum20}). To explore this curvature the spectrum was fitted in the
0.1--100 GeV energy band with two alternative spectral models with respect to the
simple power-law: a broken power-law (BPL),  $dN/dE$ $\propto$
$(E/E_{0})^{-\Gamma_1}$ for $E < E_b$,  $dN/dE$ $\propto$
$(E/E_{0})^{-\Gamma_2}$ for $E > E_b$, where $\Gamma_1$ and $\Gamma_2$ are the
photon indices below and above the energy break $E_b$; a LogParabola (LP), $dN/dE$ $\propto$
$(E/E_{0})^{-\alpha-\beta\,log(E/E_{0})}$, where the parameter $\alpha$ is the
spectral slope at a reference energy $E_0$ and $\beta$ measures the curvature around the peak \citep[see][]{landau86, massaro04}.
In order to estimate as accurately as possible the energy break parameter of
the BPL function, we studied the profile of the likelihood function, fixing
$E_{b}$ at different values between 100 MeV and 5 GeV with a step of 50 MeV. This
likelihood profile was fitted with a parabolic function; the minimum for the parabola corresponds to $E_b$ =
(2290 $\pm$ 390) MeV.
Fixing $E_b$ at this value during the BPL likelihood fit, the results are:
prefactor = (3.31 $\pm$ 0.13)$\times$10$^{-12}$, $\Gamma_1$ = 2.10 $\pm$ 0.02,
$\Gamma_2$ = 2.69 $\pm$ 0.07, with TS = $ 15733$ and an integral flux in the
0.1--100 GeV energy range of (37.6 $\pm$ 0.8)$\times$10$^{-8}$ ph cm$^{-2}$ s$^{-1}$.
The LogParabola fit was performed fixing the reference energy $E_0$ at 300 MeV; the
normalization was found to be (4.28 $\pm$ 0.08)$\times$10$^{-10}$, $\alpha$ =
2.01 $\pm$ 0.03 and $\beta$ = 0.09 $\pm$ 0.01 with TS = $15732$ and an integral flux in the
0.1--100 GeV energy range of (38.7 $\pm$ 0.7)$\times$10$^{-8}$ ph cm$^{-2}$ s$^{-1}$. We used a
likelihood ratio test to check the PL model (null hypothesis) against the BPL model (alternative hypothesis). These values may be compared,
following \citet{nolan12}, by defining the significance of the curvature TS$_{\rm curve}$=TS$_{\rm BPL}$--TS$_{\rm PL}$=72 corresponding to
$\sim$8.5 $\sigma$. In the same way for the LP model we obtain
TS$_{\rm curve}$=71 corresponding to $\sim$8.4 $\sigma$. This shows
that the 20-month LAT spectrum of PKS 0537--441 does not follow a power-law function, and the curvature is significant.

\begin{table}
\footnotesize
\caption{Results of the spectral fits to the {\it Fermi}-LAT data in the 0.1--100 GeV energy range for the 20-month and the 5 sub-periods with a
  power-law model.}
\begin{center}
\label{table:latresult}
\begin{tabular}{lccc}
\tableline
\tableline
Period & Flux (E $>$ 100 MeV)  & $\Gamma$  & TS\\
       & [$\times$10$^{-8}$ ph cm$^{-2}$ s$^{-1}$] & & \\  
\tableline
20-month & 39.9 $\pm$ 0.8 & 2.20 $\pm$ 0.01 & 15661 \\
Period 1 & 29.5 $\pm$ 1.7 & 2.31 $\pm$ 0.04 & 899 \\
Period 2 & 60.9 $\pm$ 0.2 & 2.27 $\pm$ 0.02 & 2247 \\
Period 3 & 33.0 $\pm$ 0.4 & 2.36 $\pm$ 0.01 & 1195 \\
Period 4 & 39.7 $\pm$ 3.1 & 2.55 $\pm$ 0.08 & 2514 \\
Period 5 & 73.1 $\pm$ 2.6 & 2.09 $\pm$ 0.03 & 4500 \\
\tableline
\tableline
\end{tabular}
\end{center}
\end{table}

\begin{figure}
\centering
\includegraphics[width=8.25cm]{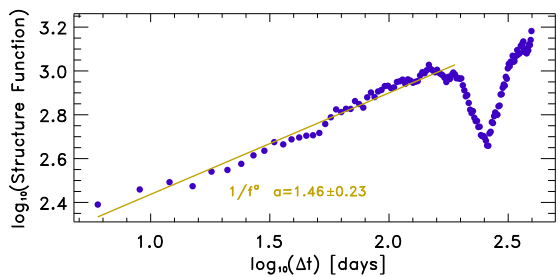}
\caption{First order structure function of the 3-day bin $\gamma$-ray light curve for the period 2008 August 4 -- 2010 April 4.} \label{figure_sf}
\end{figure}

\subsection{Temporal behaviour}

To investigate the temporal activity of PKS 0537--441, we used the light curve
of the integrated $\gamma$-ray flux over the entire period 2008 August 4--2010 April 4 with 3-day time bins in the band 0.1--100 GeV
already presented in Sect 2.

\par In Fig.~\ref{figure_sf} the first order structure function (SF) of the $\gamma$-ray light curve is reported. The SF analysis shows a power-law increasing trend from zero-lag
to ($147\pm3$) days lag, resulting in a $1/f^{a}$ power density spectrum (PDS)
with $a= 1.5\pm 0.2$. This trend implies universality from 3-day to about
5-month time-scales, and the index value represents a temporal variability
placed halfway between the flickering (red noise) and the shot noise
(Brownian-driven process) behaviour. This slope value is in agreement with the
average value of the PDS directly evaluated in the frequency domain using the 3-day bin light curves extracted for the 9 brightest FSRQs of the first year of {\em Fermi} observations \citep{Abdo2010c}.
\par
\begin{figure}
\centering
\includegraphics[width=8.25cm]{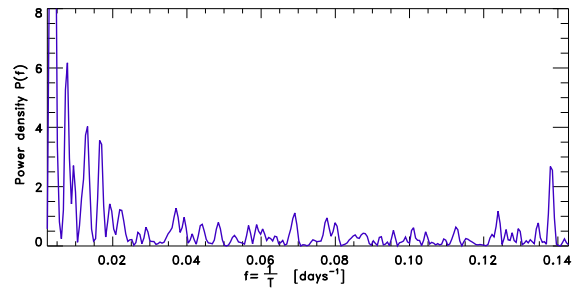}
\caption{Periodogram P(f) of the 3-day bin $\gamma$-ray
  light curve. No significant peak is present in P(f); therefore no significant power spectrum component and no periodicity is present.}
\label{FIG:periodogram}
\end{figure}

\begin{figure}
\centering
\includegraphics[width=8.25cm]{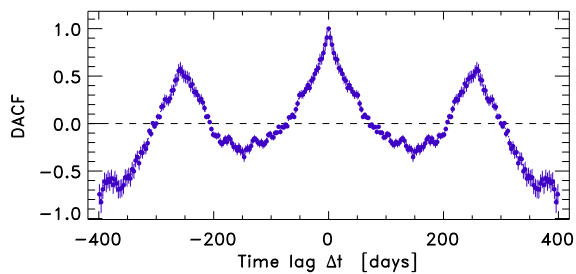}
\caption{The Discrete Autocorrelation Function applied to the
  3-day bin $\gamma$-ray light curve. DACF points out a zero-value crossing
  time lag between 69 and 72 days.}
\label{FIG:DACF}
\end{figure}

 A consistent drop is seen in Fig.~\ref{figure_sf} at lags longer than about 200 days, probably
due to both the contamination of finite-series edge effects and to the main
outburst and active phase of PKS 0537--441, which can be seen in the central part of the light curve from about MJD 54885 to about 55087. A
hint of a possible moderate break is found at timescales between 48 and 51
days, but this does not necessarily imply a local characteristic time scale,
considering that the periodogram of the light curve does not show
significant evidence for any characteristic and recurrent time-scale
(Fig.~\ref{FIG:periodogram}). The discrete autocorrelation function (DACF) does not show peaks at
time lags below the 200-day limit, and the zero-crossing time lag is placed between 69 and 72 days (Fig.~\ref{FIG:DACF}).

\begin{figure*}
\centering
\includegraphics[width=5cm]{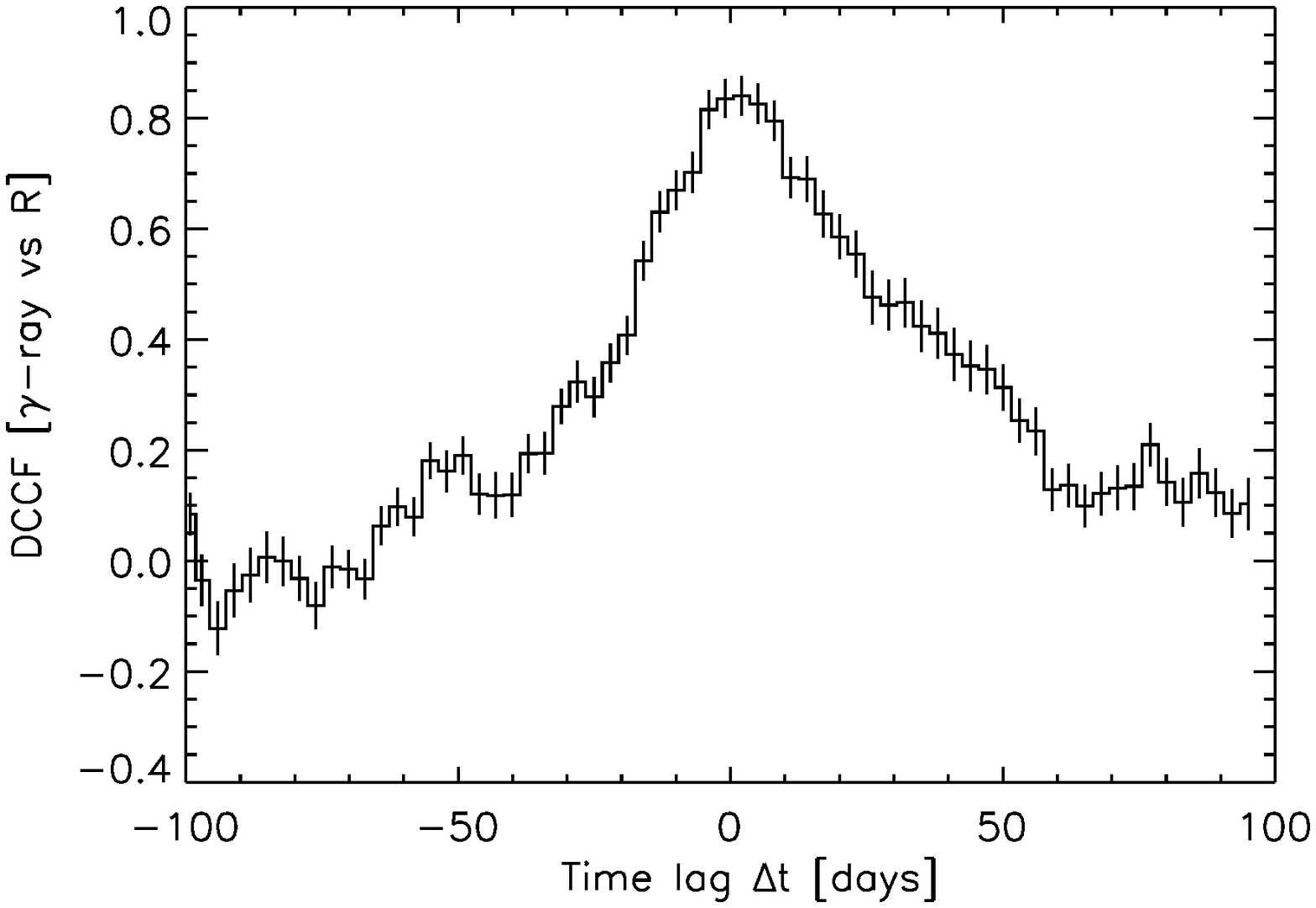}
\hspace{3mm}
\includegraphics[width=5cm]{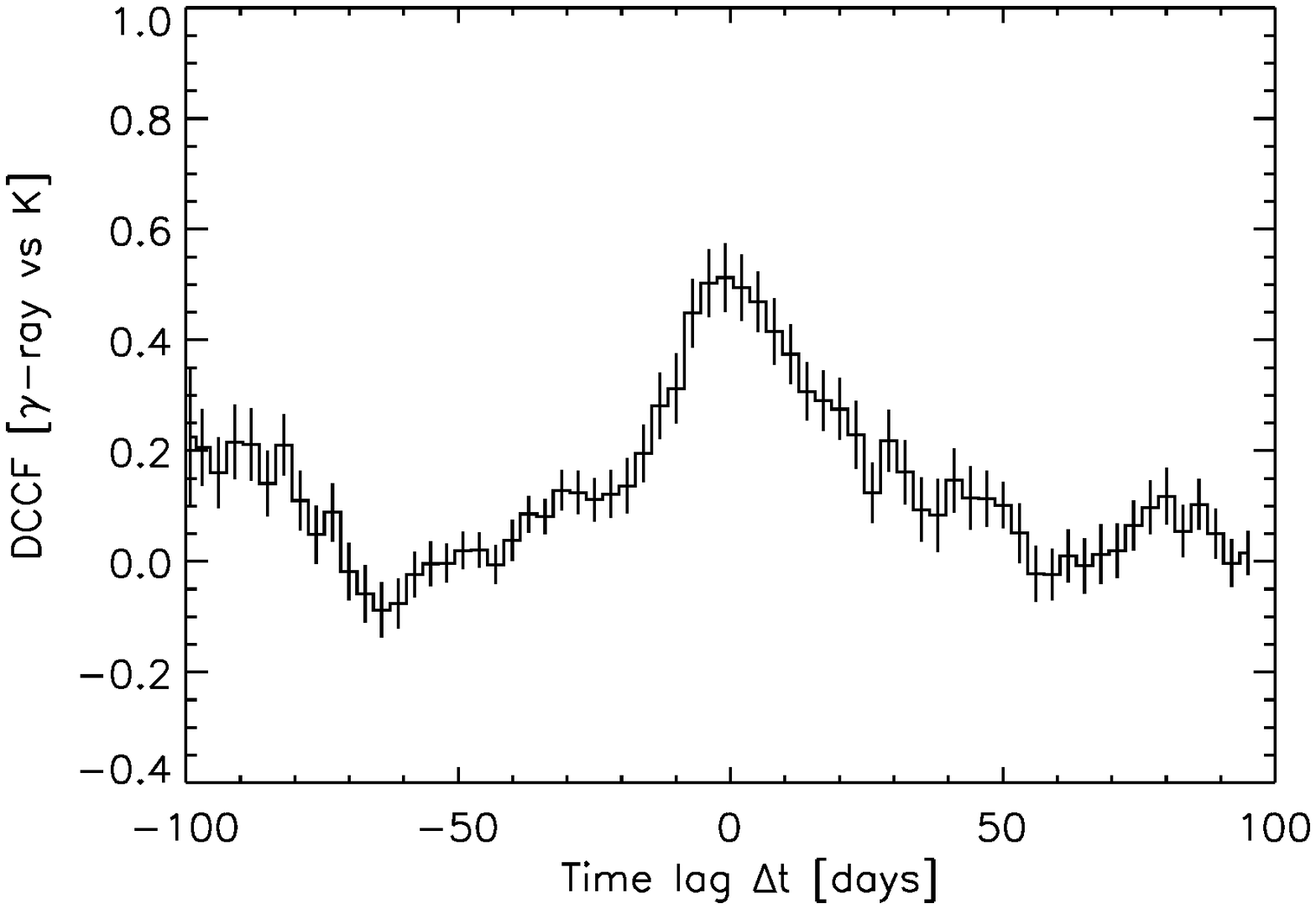}
\hspace{3mm}
\includegraphics[width=5cm]{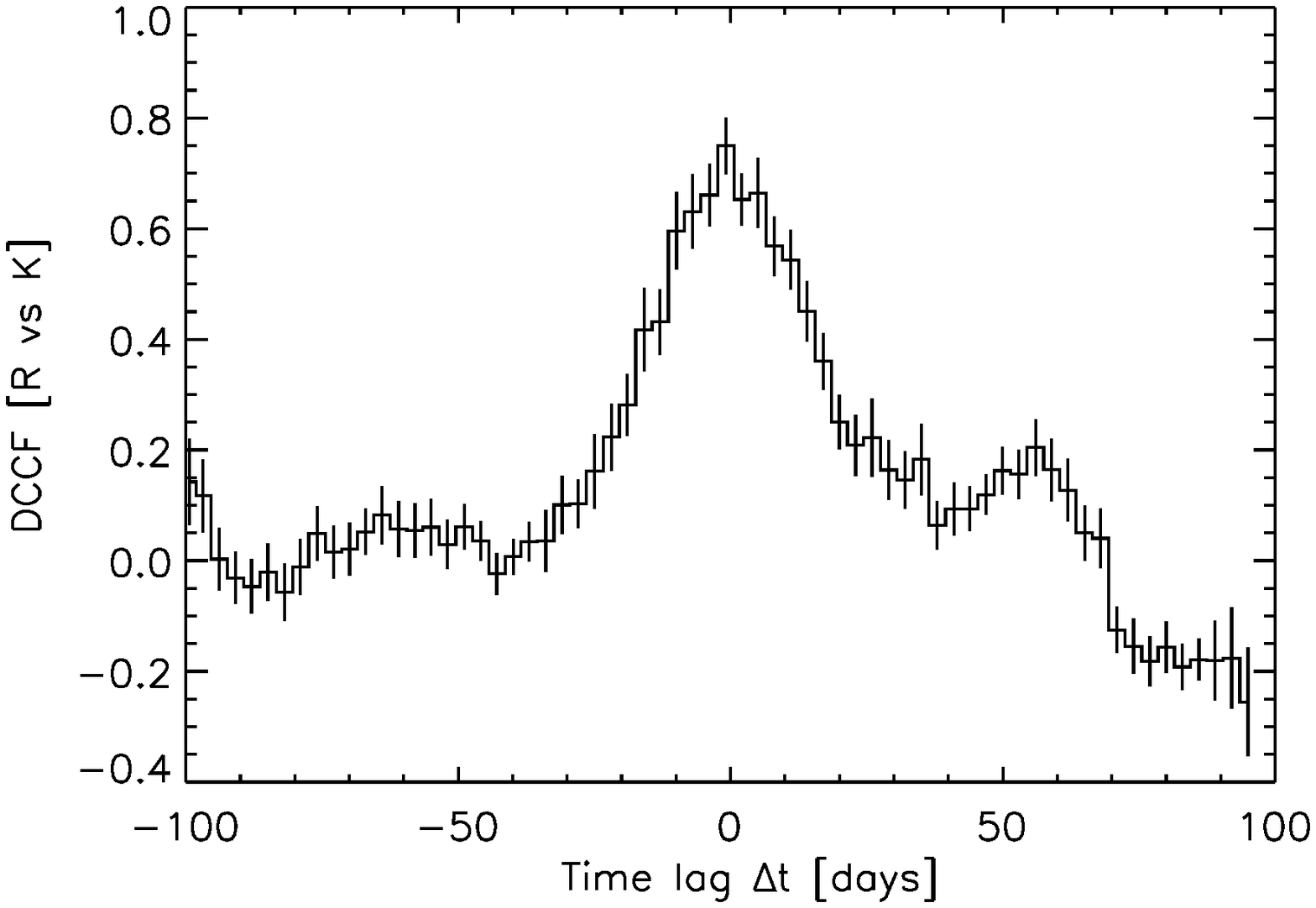}
\caption{Discrete Cross Correlation Function between the $\gamma$-ray and
  $R$-band (left panel), $\gamma$-ray and $K$-band (center panel), $R$-band
  and $K$-band (right panel) light curves. In each case the DCCF shows a correlation peak consistent with
  lag = 0.}
\label{DCF}
\end{figure*}

\section{Discrete Cross Correlation Analysis}

Correlations between the best sampled multiwavelength light curves of PKS
0537$-$441 were investigated by computing the discrete cross correlation
function (DCCF), following \citet{edelson88}. Lag and correlation strength were
computed by fitting a gaussian profile to the correlation peak in
the DCCF and uncertainties in these parameters were estimated by a
Monte Carlo method \citep[see][]{peterson98}. DCCFs were
calculated both with detrended and non-detrended light curves \citep[see
  e.g.][for details about the detrending procedure]{welsh99}  in
order to look for differences in correlation of the rapid
variations and the slower long term variability.

The combined ATOM and REM $R$-band data provide the best overlap with the
LAT $\gamma$-ray light curve. The DCCF between these two data sets, without
detrending, is shown in Fig.~\ref{DCF}, left panel (here and in the following positive lag
in $\gamma$ rays - R means that $\gamma$-ray flux variations lead those in the $R$-band).
The peak in Fig.~\ref{DCF} (left panel) corresponds to a correlation of 0.92 $\pm$ 0.06 when
corrected for the effect of measurement errors (which reduces the
correlation). A Spearman rank correlation test between the $\gamma$-ray and
$R$-band flux gives coefficient r$_{s}$ of 0.6609 with a probability of chance
occurence $< 10^{-6}$, confirming the positive correlation.
The correlation between the $\gamma$-ray and the $K$-band is weaker, 0.54 $\pm$ 0.07
(Fig.~\ref{DCF}, center panel). This is not an effect of the sparser sampling
of the $K$-band light curve, since the difference is present also when the analysis is limited to the time
range where both bands are well sampled (MJD 54890--55290). Furthermore the DCCFs show no significant time lag, with an uncertainty of a few days, between the $R$ and
$K$ bands (Fig.~\ref{DCF}, right panel).

\begin{table}
\footnotesize
\caption{Time lag and strength of the cross correlation peak estimated by a
  gaussian fit.  All data were used in the case with no detrending, while the
  detrended light curves used data from MJD 54890 to 55290. The correlation
  values have been corrected for the effect of white noise due to measurement errors.}
\begin{center}
\label{corr}
\begin{tabular}{lcccc}
\tableline 
\tableline 
& \multicolumn{2}{c}{Not detrended} & \multicolumn{2}{c}{Detrended}\\
DCCF & Lag (days) & Max & Lag (days) & Max\\
\tableline
$\gamma$ rays - R & 3.1 $\pm$ 1.9 & 0.92 $\pm$ 0.06 & 0.5 $\pm$ 2.4 & 0.57 $\pm$ 0.07\\
$\gamma$ rays - K & 1.9 $\pm$ 2.5 & 0.54 $\pm$ 0.07 & 0.1 $\pm$ 2.6 & 0.43 $\pm$ 0.08\\
R - K & 0.0 $\pm$ 1.1 & 0.71 $\pm$ 0.06 & -0.5 $\pm$ 0.9 & 0.64 $\pm$ 0.05\\
\tableline
\end{tabular}
\end{center}
\end{table}

\begin{figure}
\centering
\includegraphics[width=8.25cm]{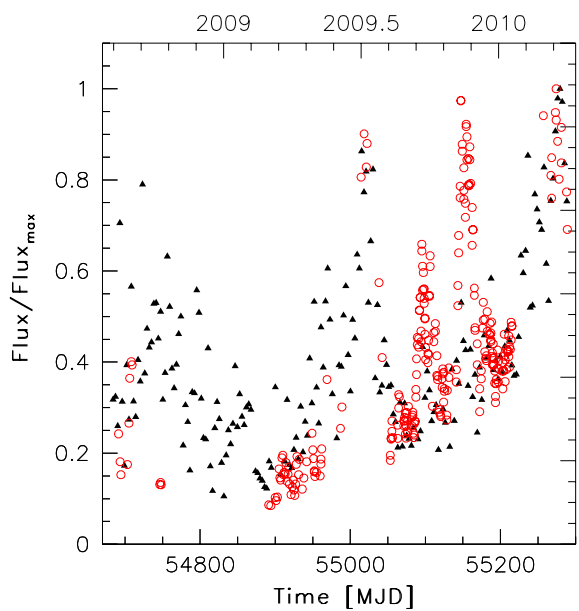}
\caption{Comparison between $\gamma$-ray and $K$-band light curves. We
  superimpose $\gamma$-ray
(black triangles) and $K$-band (red empty circles) light curves normalizing $\gamma$-ray and $K$
flux values over the whole observing period to the respective peak flux values. The time binning
is 3 days for $\gamma$-ray data and 1-day for $K$-band data set.
}
\label{comparison1}
\end{figure}

\begin{figure}
\centering
\includegraphics[width=8.25cm]{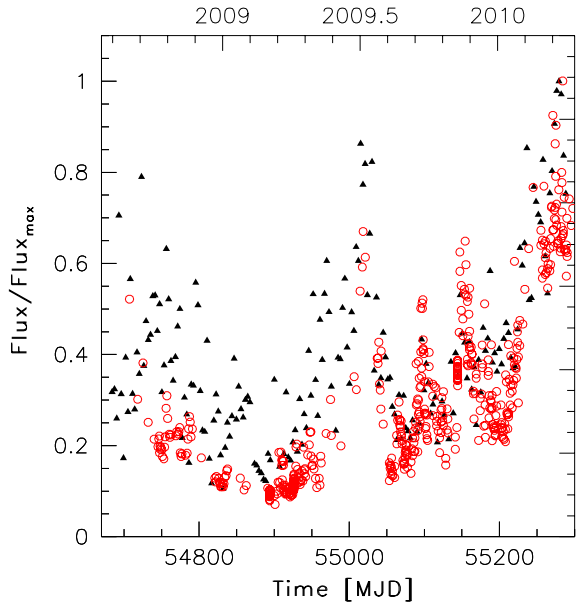}
\caption{Comparison between $\gamma$-ray and $R$-band light curves. We
  superimpose $\gamma$-ray
(black triangles) and $R$-band (red empty circles) REM light curves normalizing $\gamma$-ray and $R$
flux values over the whole observing period to the respective peak flux values. The time binning
is 3-day for $\gamma$-ray data and 1-day for $R$-band data set.
}
\label{comparison2}
\end{figure}

It is clear from the light curves that there is a correlation on long time
scales and this is also shown by the results of the DCCFs for the detrended
light curves (see Table.~\ref{corr}). The correlations are weaker compared to the non-detrended ones
and also here no significant time lag was detected. The position of the peak
is consistent with zero time lag (1.3 $\pm$ 1.5 days). When both light curves were detrended with a
second order polynomial the correlation was reduced to 0.51 $\pm$ 0.07, but
still consistent with zero time lag (0.3 $\pm$ 1.8 days). However different
behaviours are quite evident going into details of the NIR, optical, and
$\gamma$-ray light curves (see Figs.~\ref{comparison1} and ~\ref{comparison2}). A similar situation has
been already observed for other bright $\gamma$-ray blazars \citep[e.g. 3C
  454.3, 4C $+$38.41;][]{raiteri11, raiteri12}. This complex behaviour could be in agreement with the turbulent extreme multi-cells
scenario proposed by \citet{marscher12}. In this context the slope of the PDS
obtained for the $\gamma$-ray light curve indicates a significant contribution
of red noise and thus of a random process such as the turbulence that is modulating the emission. 

\section{X-ray to mm behaviour}

The BAT spectrum of the data collected during 2004 November--2009
August was fit with a power law ($dN/dE\propto
E^{-\Gamma_{\rm BAT}}$) with a photon index $\Gamma_{\rm BAT}$ =
1.5 $\pm$ 0.5 and a 15--150 keV flux of (1.5$^{+0.7}_{-0.5})$ $\times$
$10^{-11}$ erg cm$^{-2}$ s$^{-1}$ ($\chi^2_{\rm red}$ = 0.968, with 6
d.o.f.). As a comparison the same photon index with a flux a factor of
two higher was obtained for the BAT spectrum extracted from 2004
December to 2005 November, dominated by episodes of more intense
activity \citep{pian07}, indicating a quite stable spectral shape in hard X-rays.

During 2008 October -- 2010 March {\em Swift}/XRT observed the source
with a 0.3--10 keV flux in the range (3.2--7.7) $\times$ 10$^{-12}$
erg cm$^{-2}$ s$^{-1}$, with a photon index varying in the range
1.6--1.9. The interval of photon index values is very similar to that
previously observed by {\em Swift}/XRT spanning a larger flux range \citep[$\Gamma_{\rm X}$ = 1.6--1.8,][]{pian07}. A Spearman rank correlation test between the $\gamma$-ray and the X-ray flux
gives a coefficient r$_{s}$ of 0.6176 with a probability of chance occurrence of
8$\times$10$^{-3}$, indicating a positive correlation.
The small
variability amplitude (calculated as the ratio of maximum to minimum flux) observed in X-rays ($\sim$ 2.5) with respect to the
$\gamma$ rays ($\sim$ 11) could be an indication that the X-ray
emission is produced by the low-energy tail of the same electron
distribution. The peak flux observed in 2008 October is a factor of
two lower with respect to the 2004--2005 {\em Swift}/XRT observations
\citep{pian07}. Interestingly the high X-ray flux in 2008 October
coincides with a low activity state both in NIR-to-UV and $\gamma$-ray
bands, suggesting that a second emission component can give a
significant contribution in soft X-rays in that period (e.g.~an
increase of the synchrotron self-Compton component). We noted that
X-ray flux variations had no optical/UV counterpart also in 2005
November \citep{pian07}. Similarly an X-ray flare with no
  counterpart at other frequencies has been observed in 3C 279
  \citep{abdo10, hayashida12}.

\begin{figure}
\centering
\includegraphics[width=8.0cm]{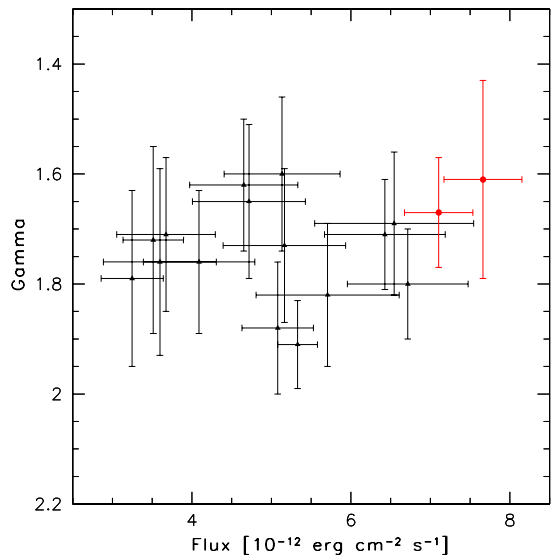}
\caption{{\em Swift}/XRT photon index of PKS 0537$-$441 as a function of the
  0.3--10 keV flux. The observations performed on 2008 October are shown with red circles.}
\label{XRT_hwb}
\end{figure}

Fig.~\ref{XRT_hwb} shows the XRT photon indices as a
function of the fluxes in the 0.3--10 keV band; no significant change
in the X-ray spectrum with the increase in the flux has been observed,
and thus there is no hint of an hard-spectrum additional component becoming
persistent and significant at higher fluxes. However we note a relatively hard spectrum ($\Gamma \sim$1.6-1.7) during the 2008 October observations (red circles in Fig.~\ref{XRT_hwb}). The high
correlation between the X-ray and optical ($V$-band) light curves
reported in \citet{pian07} is not completely confirmed by our
long-term monitoring of the source. In fact the peak of the activity
in $V$-band observed by {\em Swift}/UVOT in 2010 March has no
counterpart in the X-ray light curve. This, together with the increase
of the flux observed in X-rays in October 2008 with no counterpart at
other wavelengths, suggests a more complex connection between optical,
X-ray (and $\gamma$ ray) emission over a long period. However the
sparse sampling of the X-ray light curve does not allow us to make a precise
comparison with the NIR, optical as well as $\gamma$-ray light
curves. By contrast, the $K$-band and $R$-band are well sampled and a
  comparison with the $\gamma$-ray light curve showed an overall correlation
  (see Sect.~5). However on inspecting the light curves on short time scales
  we note a significant increase of the NIR and optical
  fluxes in 2009 September and November with a smaller
  variability amplitude in $\gamma$ rays (see Figs.~\ref{comparison1} and
  \ref{comparison2}), peaking on 2009 September 21 (MJD 55095) and November
  12 (MJD 55147) (see Fig.~\ref{FIG:MW_lc}). The largest amplitude variations
  are usually detected in $\gamma$ rays \citep{Abdo2010c}, and thus this
  behaviour is quite peculiar. Recently, a different behaviour between $\gamma$-ray
and NIR/optical light curves has been observed in the intermediate-frequency-peaked BL Lac 3C 66A in 2008 and 2009--2010, with a good
correlation in 2008 and an increasing NIR/optical flux with no counterpart in
$\gamma$ rays during 2009--2010 \citep{itoh12}. In the same way,
notwithstanding a significant correlation estimated between optical and
$\gamma$-ray bands during 2008--2011 a strong optical flare without a
$\gamma$-ray counterpart has been detected from the FSRQ 4C $+$38.41 in 2011 July \citep{raiteri12}.

During 2008--2010 an increase with an amplitude of a factor of $\sim$7
was observed in $V$-band, significantly lower with respect to
2004--2005 observations (a factor of $\sim$60), confirming the huge
flaring activity observed by REM and {\em Swift} in 2005
\citep{dolcini05, pian07}. A decreasing variability amplitude was
observed from NIR (a factor of $\sim$12--13) to UV (a factor of
$\sim$5). In particular a clear change of variability amplitude was
observed between $R$-band (a factor of $\sim$14) and $V$-band (a
factor of $\sim$7).

There appears to be a good correlation between the 230 GHz light curve
collected by SMA and the $\gamma$-ray light curve, in particular a similar increase was
observed in 2010 reaching a peak value of 8.05 Jy on 2010 March 9 for
SMA during the brightest $\gamma$-ray flaring period. The lack of SMA
observations after this date does not allow us to determine if the 230
GHz flux density is still increasing and whether the mm-peak is
strictly simultaneous with the $\gamma$-ray peak. However the
contemporaneous increase in the two energy bands in 2010 seems to
indicate that the emission region at mm and $\gamma$ rays is
co-spatial and thus the $\gamma$-ray flaring activity probably 
originates at large distance from the central engine, beyond the broad line
region (BLR).

\section{Modeling the SEDs}

\begin{figure}
\centering
\includegraphics[width=8.0cm]{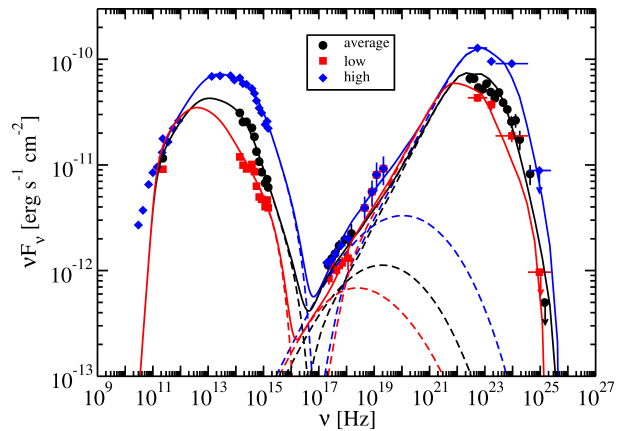}
\caption{Modeling of the SEDs of PKS 0537$-$441 in a low, average,
  and high state including {\em Fermi}, {\em Swift}, ATOM,
REM, and SMA data. See text for details.}
\label{SED}
\end{figure}

We build three SEDs for PKS 0537$-$441 in three different activity states: low, average, and high. We used REM, ATOM, and {\em Swift} data collected on 2009
March 4 (MJD 54894), 2009 June 4 (MJD 54986), and 2010 March (MJD 55258), and LAT spectra calculated over the period 2009 February 28
-- May 11 (MJD 54890--54962), 2008 August 4 -- 2010 February 4 (MJD 54682--55231), and 2010
February 28 -- April 4 (MJD 55255--55290) for the low, average, and high activity states, respectively. For each SED we used the SMA observation available nearest to
the {\em Swift} observation, i.e. 2009 February 27, 2009 July 15, and 2010
March 7. For the high activity state also the {\em Planck} and {\em Wide-Field Infrared Survey Explorer (WISE)} data
collected on 2010 March 7--22 \citep[from][]{giommi12} and March 4--7, respectively, has been reported. Finally we report in the SED the average BAT spectrum.

All three SEDs indicate a lower-frequency component peaking at $\sim
10^{13}-10^{14}$\ Hz and a higher-frequency component peaking around $\sim
10^{21}-10^{22}$\ Hz. We attempted to reproduce the SEDs with
leptonic models which include synchrotron and synchrotron self-Compton
(SSC) scattering \citep{finke08}. These attempts failed, as expected
for FSRQs and some LSP BL Lacs.  We then added an external Compton
(EC) component of seed photons from a dust torus.  We were able to
obtain reasonable fits to all three states, with the only difference
between states being the electron distribution. These model fits are
shown as curves in Figure \ref{SED}, and the model parameters can be
found in Table \ref{table_fit}. See \citet{dermer09} for a
description of the model parameters. The dust torus was modelled as a
one-dimensional ring around the black hole, aligned orthogonal to
the jet, and its parameters are approximately in agreement with dust
emitting at the sublimation radius with the formula given by
\citet{nenkova08}.  The dust luminosity in the model is rather low,
but note that there is no hint of dust or disc emission in the
IR-optical portion of the SED, in agreement with the results reported in \citet{impiombato11}.  For all states, we found that an
electron distribution with two power laws (a single break) was not
enough to explain the observed SED; an additional power law and break
were needed, in particular for reproducing the NIR-optical part of the spectrum, so that the electron distribution is given by
\begin{eqnarray}
N_e(\gamma^{\prime}) \propto \gamma^{\prime -p_1}\ ,\ \gamma^{\prime}_{min} < \gamma^{\prime} < \gamma^{\prime}_{brk,1} \nonumber \\ 
\ N_e(\gamma^{\prime}) \propto \gamma^{\prime -p_2}\ ,\ \gamma^{\prime}_{brk,1} < \gamma^{\prime} < \gamma^{\prime}_{brk,2}  \nonumber \\
N_e(\gamma^{\prime}) \propto \gamma^{\prime -p_3}\ ,\ \gamma^{\prime}_{brk,2} < \gamma^{\prime} < \gamma^{\prime}_{max} \nonumber
\end{eqnarray}
For the modeling, the variability time was chosen to be a bit more
than two days, consistent with the observed light curve.  This constrains the size of the emitting region, given a
Doppler factor.  In our model fits, the primary emitting region is
optically thin to synchrotron self-absorption down to $\sim
200$\ GHz, so that it can reproduce the data at higher frequencies.
Below this frequency, presumably the radio emission comes from other,
larger emitting regions \citep[e.g.,][]{konigl81}.  In our model fits,
this blob is quite far from the black hole, $\ge 1$\ pc.  If the
emitting blob takes up the entire cross section of the jet, then it
would have an opening angle of $\theta_{open}\sim 2^{\circ}$.  This is
approximately consistent with the opening angles found for other
blazars based on multi-epoch VLBI observations \citep{jorstad05}.

\begin{table*} 
\centering
\caption{Model parameters for the SEDs of PKS 0537$-$441.}
\begin{tabular}{lcccc}
\hline
\hline
Parameter &  Symbol & Low State & Average State & High State \\
\hline
Redshift & 	$z$		& 0.896 & 0.896 & 0.896	  \\
Bulk Lorentz Factor & $\Gamma$	& 50 & 50 & 50	  \\
Doppler Factor & $\delta_{\rm D}$       & 50 & 50 & 50    \\
Magnetic Field & $B$         & 0.2 & 0.2 & 0.2 G   \\
Variability Time-Scale & $t_v$       & $2\times10^5$ & $2\times10^5$ & $2\times10^5$ \\
Comoving radius of blob & $R^{'}_{b}$ & $1.6\times10^{17}$\ cm &
$1.6\times10^{17}$\ cm & $1.6\times10^{17}$\ cm \\
Jet Height & $r$ & $3.7\times10^{18}$\ cm & $3.7\times10^{18}$\ cm & $3.7\times10^{18}$\ cm \\
First Electron Spectral Index & $p_1$       & 2.0 & 2.0 & 2.0     \\
Second Electron Spectral Index  & $p_2$ & 3.1 & 3.1 & 3.1	 \\
Third Electron Spectral Index  & $p_3$ & 3.9  & 4.6 & 4.6	 \\
Minimum Electron Lorentz Factor & $\gamma^{\prime}_{min}$  & 4.0 & 1.0 & 3.0 \\
Break Electron Lorentz Factor & $\gamma^{\prime}_{brk,1}$ & $3.0\times10^2$  & $4.7\times10^2$ & $7.0\times10^2$ \\
Break Electron Lorentz Factor & $\gamma^{\prime}_{brk,2}$ & $9.0\times10^2$ & $2.6\times10^3$ & $4.2\times10^3$ \\
Maximum Electron Lorentz Factor & $\gamma^{\prime}_{max}$  & $1.3\times10^4$ & $2.3\times10^4$ & $3.0\times10^4$ \\
Dust Torus luminosity & $L_{dust}$ & $3.3\times10^{42}$\ erg s$^{-1}$ & $3.3\times10^{42}$\ erg s$^{-1}$ & $3.3\times10^{42}$\ erg s$^{-1}$ \\
Dust Torus temperature & $T_{dust}$ & $3\times10^2$\ K & $3\times10^2$\ K & $3\times10^2$\ K \\
Dust Torus radius & $R_{dust}$ & $9.4\times10^{18}$\ cm & $9.4\times10^{18}$\ cm & $9.4\times10^{18}$\ cm \\
Jet Power in Magnetic Field & $P_{j,B}$ & $1.9\times10^{46}$\ erg s$^{-1}$ & $1.9\times10^{46}$\ erg s$^{-1}$ & $1.9\times10^{46}$\ erg s$^{-1}$ \\
Jet Power in Electrons & $P_{j,e}$      & $1.2\times10^{46}$\ erg s$^{-1}$ & $1.2\times10^{46}$\ erg s$^{-1}$ & $1.3\times10^{46}$\ erg s$^{-1}$ \\
\hline
\hline
\end{tabular}
\label{table_fit}
\end{table*}

In the ``average state'', the LAT spectrum shows a clear deviation from
a single power-law (see Section 3).  This has been found for numerous
other FSRQs and LSP BL Lacs \citep{Abdo2010d}, most notably for
the extremely $\gamma$-ray bright FSRQ 3C~454.3 \citep{abdo09_3c454.3,
  ackermann10_3c454.3, abdo11_3c454.3}.  The cause of this curvature,
often characterized as a spectral break (when fit with a double
power-law) is a bit of a mystery. It has been suggested that it is
due to a feature in the electron distribution \citep{abdo09_3c454.3};
due to $\gamma\gamma$ absorption of $\gamma$-rays with He~II Ly$\alpha$
broad line photons \citep{poutanen10,stern11}; due to a combination of
Compton scattering of two seed photon sources, for example, directly
from the accretion disc, and from broad line emission
\citep{finke10_3c454.3}; or due to Klein-Nishina effects from
Compton-scattering of H~I Ly$\alpha$ photons
\citep{ackermann10_3c454.3}. In the ``high state'' SED for PKS~0537$-$441, there is clearly a spectral break in the
IR-optical spectrum, around $3\times10^{14}$\ Hz ($\sim 1\ \mu$m).
The combination of a break found in both the synchrotron and
Compton-scattered spectrum seems to indicate a common origin for the
break, most likely due to an intrinsic feature in the underlying
electron distribution.  Indeed, that is how we model it for this
source. Although the broken IR/optical spectrum in the ``high
state'', and the curved $\gamma$-ray spectrum in the ``average state'' are
not contemporaneous, there does seem to be a hint of curvature in the
``high state'' LAT spectrum, especially when the upper limit in the
highest energy bin is considered.  It is certainly possible, maybe
even likely, that the cause of the $\gamma$-ray spectral breaks in all
blazars which show them is related to features in the electron distribution,
although one should be careful not to over-generalise.  Simultaneous
IR, optical and $\gamma$-ray observations of other blazars could be very
useful for making this determination, although contamination in the IR
by a dust torus \citep[e.g.,][]{malmrose11} or another synchrotron
component \citep[e.g.,][]{ogle11} could make this difficult.

The three states differ very little in the radio and X-rays.  This
emission is caused by synchrotron and EC emission, respectively, of
the lower portion of the electron distribution, although there is some
additional SSC emission for the X-rays.  Consequently, the jet power
in electrons (see Table \ref{table_fit}) varies little between states,
since the $p_1=2.0$ portion of the electron distribution contains most
of the electrons' energy.  The model implies emission very close to
equipartition between electrons and Poynting flux for all model fits,
with a slight dominance of Poynting flux.

\section{Conclusions}

We presented multiwavelength observations of PKS 0537$-$441 during a period of
20 months (2008 August--2010 April) including {\em Fermi}, {\em Swift}, ATOM,
REM, and SMA data. Strong variability has been observed in $\gamma$ rays, with two major
flaring episodes (2009 July and 2010 March) and a harder-when-brighter
spectral behaviour, quite common for FSRQs and LSP BL Lacs, in 2010 March. The average LAT
spectrum accumulated over 20 months showed a significant curvature, well
described both by a BPL model with $E_b$ = (2290 $\pm$ 390) MeV and a
log-parabola model. 
A quite steep PDS slope ($\alpha$ = 1.5 $\pm$ 0.2) has been estimated, in agreement with those observed for the
brightest FSRQs \citep{Abdo2010c}, suggesting a random-walk underlying mechanism.

Only moderate variability has been observed in X-rays with no correlation
between flux and photon index. An increase of the flux in the X-ray band with
no counterpart at the other wavelengths has been observed in 2008 October,
suggesting a significant contribution of a second component (e.g. SSC emission)
in some high activity states. No clear correlation of the X-ray emission with the NIR or optical bands has been found. Only a correlation with the
gamma-ray band has been observed. In addition we observed correlation between the $\gamma$-ray band with the $R$-band and more weakly with
the $K$-band, with no significant time lag. However, together with this
``overall'' correlation we note that on finer timescales the light curves showed
differences. In particular, two flaring episodes in NIR and optical bands have been observed in 2009
September and even more strongly in 2009 November with no significant counterparts in $\gamma$ rays.

As seen for FSRQs and LSP BL Lacs, the SED of the source cannot be modelled by a
simple SSC model. We included an EC component of seed photons from a dust
torus. The 230 GHz light curve showed an increase simultaneously with the
$\gamma$ rays, suggesting co-spatiality of the mm and $\gamma$-ray
emission regions at large distance from the central engine and thus the dust
torus as the possible main source of seed photons. The low, average, and high activity
SEDs of the source could be fit by changing only the electron distribution
parameters. This has been observed previously for Mrk 501 \citep{petry00}, although this is a high-synchrotron-peaked BL Lac, while PKS 0537-441 is an LSP BL Lac/FSRQ. In our modeling, we have found that two breaks in the electron distribution are necessary. A spectral break
in the NIR-optical spectrum has been found, in agreement also with a significant
change of the flux variability amplitude below and above the $R$-band. This break
together with the curvature observed in $\gamma$ rays (possibly
characterized as a spectral break) seems to indicate a common
origin for them, most likely due to an intrinsic feature in the underlying electron distribution.

Broad emission lines have been observed during a low activity state of PKS
0537$-$441 \citep{Pian2005}, in contrast with the initial classification as a BL Lac
object. However, recent studies of a large sample of blazars showed that a
classification depending only on their optical/UV spectral properties is not
efficient to distinguish FSRQs and BL Lac objects \citep{Abdo2010a, ghisellini11}. Variability and spectral properties in $\gamma$ rays indicate
a FSRQ-like behaviour, in agreement also with the SED properties. The observed
isotropic $\gamma$-ray luminosity in the 0.1--100 GeV energy range is 8.2$\times$10$^{47}$
erg s$^{-1}$ for the 20-month interval considered, reaching a peak value of
2.6$\times$10$^{48}$ erg s$^{-1}$ on 3-day time scale at the end of 2010
March. These values are comparable to those of the bright FSRQs \citep{ghisellini09}. 

The multifrequency observations presented here give new clues, but also offer
new questions on the astrophysical mechanisms at work in PKS
0537$-$441. Further mm to $\gamma$-ray observations will be fundamental to
investigate in even more detail the correlations by using the long-term variability in different energy bands to achieve a complete
understanding of the fundamental processes that underlie the behaviour of this source.

\section*{Acknowledgments}

The {\em Fermi} LAT Collaboration acknowledges generous ongoing support
from a number of agencies and institutes that have supported both the
development and the operation of the LAT as well as scientific data analysis.
These include the National Aeronautics and Space Administration and the
Department of Energy in the United States, the Commissariat \`a l'Energie Atomique
and the Centre National de la Recherche Scientifique / Institut National de Physique
Nucl\'eaire et de Physique des Particules in France, the Agenzia Spaziale Italiana
and the Istituto Nazionale di Fisica Nucleare in Italy, the Ministry of Education,
Culture, Sports, Science and Technology (MEXT), High Energy Accelerator Research
Organization (KEK) and Japan Aerospace Exploration Agency (JAXA) in Japan, and
the K.~A.~Wallenberg Foundation, the Swedish Research Council and the
Swedish National Space Board in Sweden. Additional support for science analysis during the operations phase is gratefully
acknowledged from the Istituto Nazionale di Astrofisica in Italy and the Centre National d'\'Etudes Spatiales in France.
The Submillimeter Array is a joint project between the Smithsonian
Astrophysical Observatory and the Academia Sinica Institute of Astronomy
and Astrophysics and is funded by the Smithsonian Institution and the
Academia Sinica.
We thank the Swift team for making these observations possible, the
duty scientists, and science planners.
This publication makes use of data products from the Wide-field Infrared
Survey Explorer, which is a joint project of the University
of California, Los Angeles, and the Jet Propulsion Laboratory/California Institute of Technology, funded by the National Aeronautics and Space Administration. This paper is partly based on
observations obtained with Planck (http://www.esa.int/Planck), an ESA science
mission with instruments and contributions directly funded by ESA Member
States, NASA, and Canada. Some authors acknowledge financial contribution from grant PRIN-INAF-2011. We thank the anonymous referee, Y. Tanaka, C. Dermer, S. Digel, and E. Charles for useful comments and suggestions.



\end{document}